\documentclass[prl,twocolumn,superscriptaddress]{revtex4-1}
\usepackage[T1]{fontenc}
\usepackage{times}
\usepackage{graphicx,color}
\usepackage{amsfonts,amsmath,amssymb,amsbsy}
\usepackage[colorlinks=true, linkcolor=blue, citecolor=blue, urlcolor=blue]{hyperref}
\def\emph#1{\textcolor{black}{#1}}

\begin{document}

\title{Current-Induced Dynamics and Chaos of Antiferromagnetic Bimerons}

\author{Laichuan Shen}
\thanks{These authors contributed equally to this work.}
\affiliation{School of Science and Engineering, The Chinese University of Hong Kong, Shenzhen, Guangdong 518172, China}
\affiliation{College of Physics and Electronic Engineering, Sichuan Normal University, Chengdu 610068, China}

\author{Jing Xia}
\thanks{These authors contributed equally to this work.}
\affiliation{School of Science and Engineering, The Chinese University of Hong Kong, Shenzhen, Guangdong 518172, China}

\author{Xichao Zhang}
\affiliation{School of Science and Engineering, The Chinese University of Hong Kong, Shenzhen, Guangdong 518172, China}

\author{Motohiko Ezawa}
\affiliation{Department of Applied Physics, The University of Tokyo, 7-3-1 Hongo, Tokyo 113-8656, Japan}

\author{Oleg A. Tretiakov}
\affiliation{School of Physics, The University of New South Wales, Sydney 2052, Australia}

\author{Xiaoxi Liu}
\affiliation{Department of Electrical and Computer Engineering, Shinshu University, 4-17-1 Wakasato, Nagano 380-8553, Japan}

\author{Guoping Zhao}
\email[Email:~]{zhaogp@uestc.edu.cn}
\affiliation{College of Physics and Electronic Engineering, Sichuan Normal University, Chengdu 610068, China}

\author{Yan Zhou}
\email[Email:~]{zhouyan@cuhk.edu.cn}
\affiliation{School of Science and Engineering, The Chinese University of Hong Kong, Shenzhen, Guangdong 518172, China}

\begin{abstract}
A magnetic bimeron is a topologically non-trivial spin texture carrying an integer topological charge, which can be regarded as the counterpart of skyrmion in easy-plane magnets. The controllable creation and manipulation of bimerons are crucial for practical applications based on topological spin textures. Here, we analytically and numerically study the dynamics of an antiferromagnetic bimeron driven by a spin current. Numerical simulations demonstrate that the spin current can create an isolated bimeron in the antiferromagnetic thin film via the damping-like spin torque. The spin current can also effectively drive the antiferromagnetic bimeron without a transverse drift. The steady motion of an antiferromagnetic bimeron is analytically derived and is in good agreement with the simulation results. Also, we find that the alternating-current-induced motion of the antiferromagnetic bimeron can be described by the Duffing equation due to the presence of the nonlinear boundary-induced force. The associated chaotic behavior of the bimeron is analyzed in terms of the Lyapunov exponents. Our results demonstrate the inertial dynamics of an antiferromagnetic bimeron, and may provide useful guidelines for building future bimeron-based spintronic devices.
\end{abstract}

\date{December 24, 2019}

\maketitle


\textit{Introduction.}~Topologically protected magnetic textures, such as magnetic skyrmions~\cite{Roszler_NATURE2006,Nagaosa_NNANO2013,Finocchio_JPD2016,Kang_PIEEE2016,Fert_NATREVMAT2017,ES_JAP2018,Zhou_NSR2018}, have attracted a lot of attention, because they have small size and can be used as non-volatile information carriers in future spintronic devices~\cite{Zhang_SciRep2015,Prychynenko_PRApplied2018,Nozaki_APL2019}.
The existence of magnetic skyrmions has been experimentally confirmed in many systems with bulk or interfacial Dzyaloshinskii-Moriya interaction (DMI)~\cite{Nagaosa_NNANO2013,Finocchio_JPD2016,Fert_NATREVMAT2017,ES_JAP2018}.
In addition, various topological structures, such as antiferromagnetic (AFM) skyrmions~\cite{Zhang_SREP2016A,Barker_PRL2016}, ferrimagnetic skyrmions~\cite{Woo_NATCOM2018}, antiskyrmions~\cite{Nayak_Nature2017}, biskyrmions~\cite{Yu_NATCOM2014}, bobbers~\cite{Rybakov_PRL2015}, and bimerons~\cite{Ezawa_PRB2011,Zhang_SciRep2015,Lin_PRB2015,Heo_SciRep2016,Leonov_PRB2017,Kharkov_PRL2017,Kolesnikov_SciRep2018,Chmiel_NatMater2018,Yu_Nat2018,Woo_Nat2018,Gobel_PRB2019,Fernandes_SSC2019,Murooka2019,Kim2019,Gao_NatCom2019}, are also current hot topics.
In particular, a bimeron consists of two merons, which can be found in easy-plane magnets~\cite{Zhang_SciRep2015,Lin_PRB2015,Leonov_PRB2017,Murooka2019}, frustrated magnets~\cite{Kharkov_PRL2017}, and magnets with anisotropic DMI~\cite{Gobel_PRB2019}.
The bimeron is a localized spin texture similar to magnetic skyrmion, which can be constructed by rotating the spin direction of a skyrmion by $90^\circ$. Magnetic bimerons can also be used as information carriers for spintronic devices made of in-plane magnetized thin films~\cite{Zhang_SciRep2015,Gobel_PRB2019,Murooka2019}.

On the other hand, AFM materials are promising for building advanced spintronic devices due to their zero stray fields and ultrafast spin dynamics~\cite{Baltz_RMP2018,Jungwirth_NNANO2016,Smejkal_NATP2018}.
Several theoretical studies~\cite{Zhang_SREP2016A,Barker_PRL2016,Khoshlahni_PRB2019,Yang_PRL2018} predict that skyrmions may exist in AFM systems, which can be manipulated by spin currents~\cite{Zhang_SREP2016A,Barker_PRL2016} and magnetic fields~\cite{Khoshlahni_PRB2019}.
Compared to ferromagnetic (FM) skyrmions, AFM skyrmions do not show the skyrmion Hall effect~\cite{Jiang_NatPhys2017,Litzius_NatPhys2017} due to zero net Magnus force, so that they can move perfectly along the driving force direction with ultrahigh speed~\cite{Zhang_SREP2016A,Barker_PRL2016,Shen_PRB2018,Zhang_NATCOM2016}.
Various methods have been proposed to control the AFM textures, such as by using spin currents~\cite{Hals_PRL2011,Shiino_PRL2016,Velkov_NJP2016}, magnetic anisotropy gradients~\cite{Shen_PRB2018}, temperature gradients~\cite{Selzer_PRL2016,Khoshlahni_PRB2019}, and spin waves~\cite{Qaiumzadeh_PRB2018}. 

For AFM systems, the motion equation of the AFM order parameter (N{\'e}el vector) is related to the second derivative with respect to time, whereas the FM Landau-Lifshitz-Gilbert (LLG) equation~\cite{Gilbert_IEEE2004} is of the first order~\cite{Baltz_RMP2018}.
Therefore, the dynamics of AFM spin textures are different from that of FM spin textures.
For example, the oscillation frequency of AFM skyrmion-based spin torque nano-oscillators (STNOs) is higher than that of FM skyrmion-based STNOs as AFM skyrmions obey the inertial dynamics~\cite{Shen_APL2019}.
In addition, the motion equation of the systems, such as the LLG equation, is usually nonlinear, resulting in the dynamic behavior being complex or even chaotic.~\cite{Moon_SciRep2014,Yang_PRL2007}
Note that for the chaos, the nonlinearity is a necessary condition rather than a sufficient condition, so that not all nonlinear systems will exhibit the chaotic behavior. In nanoscale spintronic devices, spin torque nano-oscillators are interesting candidates for chaotic systems~\cite{Devolder_PRL2019,Matsumoto_PRApplied2019,Petit-Watelot_NatPhys2012}, which are promising for various applications~\cite{Fukushima_APE2014,Ditto_Chaos2015,Wang_IEEE1999}.
For the AFM bimeron, however, its dynamics induced by a spin current still remain elusive.

In this Letter, we report the dynamics of an AFM bimeron induced by the spin current. Our theoretical and numerical results show that an isolated bimeron can be created and driven in the AFM thin film by spin currents. Furthermore, when an alternating current is applied to drive the AFM bimeron, the motion of the bimeron in a nanodisk can be described by the Duffing equation, which describes the oscillation of an object with a mass under the action of nonlinear restoring forces.
The chaotic behavior associated is also analyzed in terms of the Lyapunov exponents.

\textit{Model and theory.}~We consider a G-type AFM film with sublattice magnetization $\boldsymbol{M}_{\text{1}}$ and $\boldsymbol{M}_{\text{2}}$. By linearly combining the reduced magnetizations $\boldsymbol{m}_{\text{1}}$ and $\boldsymbol{m}_{\text{2}}$ ($\boldsymbol{m}_{i}=\boldsymbol{M}_{i}/M_\text{S}$ with the saturation magnetization $M_\text{S}$), we obtain the staggered magnetization (or N{\'e}el vector) $\boldsymbol{n}=(\boldsymbol{m}_{\text{1}}-\boldsymbol{m}_{\text{2}})/2$ and the total magnetization $\boldsymbol{m}=(\boldsymbol{m}_{\text{1}}+\boldsymbol{m}_{\text{2}})/2$, where the former could be used to describe the AFM order, while the latter is related to the canting of magnetic moments. Here we are interested in most realistic cases where the AFM exchange interaction is significantly strong, so that $\boldsymbol{m}^{\text{2}} \ll \boldsymbol{n}^{\text{2}} \sim {1}$~\cite{Dasgupta_PRB2017,Zarzuela_PRB2018}. $\boldsymbol{m}$ and $\boldsymbol{n}$ obey the following two coupled equations~\cite{Hals_PRL2011,Shiino_PRL2016,Velkov_NJP2016,Tveten_PRB2016}
%
\begin{align}
\label{eq:1a}
\boldsymbol{\dot{n}}&=(\gamma\boldsymbol{f}_{2}-\alpha\boldsymbol{\dot{m}})\times\boldsymbol{n}+\boldsymbol{T}_{1,\text{SOT}}+\boldsymbol{T}_{1,\text{STT}},\tag{1a} \\
\boldsymbol{\dot{m}}&=(\gamma\boldsymbol{f}_{1}-\alpha\boldsymbol{\dot{n}})\times\boldsymbol{n}+\boldsymbol{T}_{nl}+\boldsymbol{T}_{2,\text{SOT}}+\boldsymbol{T}_{2,\text{STT}},\tag{1b}\label{eq:1b}
\end{align}
%
where $\gamma$ and $\alpha$ are the gyromagnetic ratio and the damping constant respectively, and $\boldsymbol{T}_{nl}=(\gamma\boldsymbol{f}_{2}-\alpha\boldsymbol{\dot{m}})\times\boldsymbol{m}$ is the higher-order nonlinear term~\cite{Hals_PRL2011}. 
$\boldsymbol{T}_{1,\text{SOT}}=\gamma H_{d}\boldsymbol{m}\times\boldsymbol{p}\times\boldsymbol{n}$ and $\boldsymbol{T}_{2,\text{SOT}}=\gamma H_{d}\boldsymbol{n}\times\boldsymbol{p}\times\boldsymbol{n}$ are damping-like spin-orbit torques (SOTs), where $\boldsymbol{p}$ is the polarization vector and $H_{d}$ relates to the applied current density $j$, defined as $H_{d}=j\hbar P/(2\mu_{0}eM_\text{S}t_{z})$ with the reduced Planck constant $\hbar$, the spin polarization rate $P$, the vacuum permeability constant $\mu_{0}$, the elementary charge $e$, and the layer thickness $t_{z}$. 
$\boldsymbol{T}_{1,\text{STT}}=\gamma\eta\partial_{x}\boldsymbol{n}$ and $\boldsymbol{T}_{2,\text{STT}}=\gamma\beta\partial_{x}\boldsymbol{n}\times\boldsymbol{n}$ stand for spin-transfer torques (STTs) with the adiabatic (nonadiabatic) parameter $\eta$ ($\beta$). 
In our simulations, $\eta = 0.1\beta$ and $\beta = H_{d}t_{z}$ are adopted. $\boldsymbol{f}_{1}=-\delta E/\mu_{\text{0}}M_\text{S}\delta\boldsymbol{n}$ and $\boldsymbol{f}_{2}=-\delta E/\mu_{\text{0}}M_\text{S}\delta\boldsymbol{m}$ are the effective fields.
From a classical Heisenberg Hamiltonian~\cite{Tveten_PRB2016}, the AFM energy $E$ can be written as $E = \int{\mathcal{F}} \,dV$, where $\mathcal{F}=\frac{\lambda}{2}\boldsymbol{m}^{\text{2}}+L\boldsymbol{m}\cdot(\partial_{x}\boldsymbol{n}+\partial_{y}\boldsymbol{n})+\frac{A}{2}[(\nabla\boldsymbol{n})^{\text{2}}+\partial_{x}\boldsymbol{n}\cdot\partial_{y}\boldsymbol{n}]-\frac{K}{2}(\boldsymbol{n}\cdot\boldsymbol{n_{e}})^{2}+w_{D}$ with the homogeneous exchange constant $\lambda$, parity-breaking constant $L$~\cite{Tveten_PRB2016,Shiino_PRL2016,Qaiumzadeh_PRB2018}, inhomogeneous exchange constant $A$ and magnetic anisotropy constant $K$.
$\boldsymbol{n_{e}}=\boldsymbol{e_{x}}$ stands for the direction of the anisotropy axis and $w_{D}$ is the DMI energy density, $w_{D}=\frac{D}{2}[n_{x}(\partial_{y}n_{y}-\partial_{x}n_{z})-n_{y}\partial_{y}n_{x}+n_{z}\partial_{x}n_{x}]$ with the DMI constant $D$~\cite{Gobel_PRB2019,Velkov_NJP2016,Zarzuela_PRB2018,Rohart_PRB2013}.
Such a DMI energy can stabilize the bimeron, which can be induced at the antiferromagnet/heavy metal interface~\cite{Gobel_PRB2019}. In addition, to form the bimeron, antiferromagnets with in-plane easy-axis anisotropy, such as NiO~\cite{Baltz_RMP2018}, are favorable.

\begin{figure}[t]
\centerline{\includegraphics[width=0.48\textwidth]{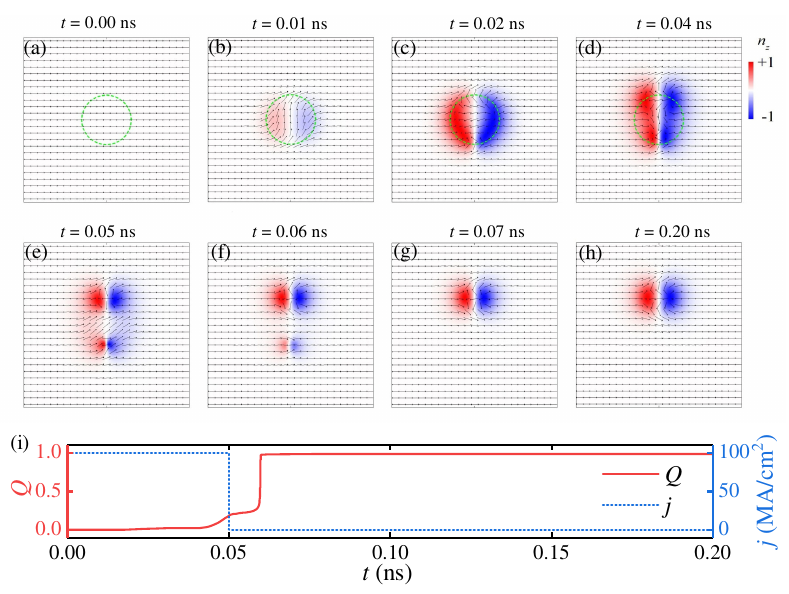}}
\caption{(a)-(h) The time evolution of the N{\'e}el vector induced by a spin-polarized current with the polarization vector $\boldsymbol{p} = -\boldsymbol{e_{z}}$, where the damping-like spin-orbit torque (SOT) is taken into account and the color represents the out-of-plane component of the N{\'e}el vector. (i) The evolution of the topological charge $Q$ and the injected current density $j$. In our simulations, the current of $j=100$ MA/cm$^{2}$ is injected in the central circular region with a diameter of $30$ nm [see green lines in Figs. (a)-(d)] and we adopt the following parameters~\cite{Barker_PRL2016}, $A=6.59$ pJ/m, $K=0.116$ MJ/m$^{3}$, $D=0.6$ mJ/m$^{2}$, $M_\text{S}=376$ kA/m, $\lambda=150.9$ MJ/m$^{3}$, $L=22.3$ mJ/m$^{2}$, $\gamma=2.211 \times 10^{5}$ m/(A s), $\alpha=0.2$ and $P=0.4$. The mesh size of $1 \times 1 \times 1$ nm$^{3}$ is used to discretize the AFM film with the size $200 \times 200 \times 1$ nm$^{3}$. Figs. (a)-(h) only show the N{\'e}el vector in the $100 \times 100$ nm$^{2}$ plane.}
\label{FIG1}
\end{figure}

Based on Eqs.~(\ref{eq:1a}) and~(\ref{eq:1b}), one can simulate the evolution of the staggered magnetization, and also derive the steady motion equations for a rigid AFM bimeron by using Thiele (or collective coordinate) approach~\cite{Thiele_PRL1973,Tveten_PRL2013,Tretiakov_PRL2008,Clarke_PRB2008} (see Ref.~\cite{Shen_SI} for details), written as
\begin{equation}
\boldsymbol{a}\cdot\boldsymbol{M}_{\text{eff}}=\boldsymbol{F}_{\alpha}+\boldsymbol{F}_{\text{SOT}}+\boldsymbol{F}_{\text{STT}},\tag{2}
\label{eq:2}
\end{equation}
where $\boldsymbol{a}$ is the acceleration, and $\boldsymbol{M}_\text{eff}$ is the effective AFM bimeron mass, which is defined as $\mu_{0}^{2}M_\text{S}^{2}t_{z}\boldsymbol{d}/\gamma^{2}\lambda$ with the dissipative tensor $\boldsymbol{d}$.
The effective AFM texture mass $\boldsymbol{M}_\text{eff}$ originates from the existence of two sublattices,~\cite{Baltz_RMP2018} and it is intrinsic.
The components $d_{ij}$ of the dissipative tensor are $d_{xx}=d_{yy}=d=\int{dxdy}(\partial_{x}\boldsymbol{n}\cdot\partial_{x}\boldsymbol{n})$ and $d_{xy}=d_{yx}=\text{0}$.
In Eq.~(\ref{eq:2}), the forces induced by the surrounding environment (e.g., the boundary effect) are not taken into account, $\boldsymbol{F}_{\alpha}=-\alpha\mu_{\text{0}}M_\text{S}t_{z}\boldsymbol{v}\cdot\boldsymbol{d}/\gamma$ represents the dissipative force with the velocity $\boldsymbol{v}$, and $\boldsymbol{F}_{\text{SOT}}$ and $\boldsymbol{F}_{\text{STT}}$ are the forces induced by SOTs and STTs, respectively.

\textit{Creation of an AFM bimeron by a spin current.}~Creating an isolated AFM bimeron is essential for practical applications. Here we employ a current to create an AFM bimeron via SOTs. As shown in Fig.~\ref{FIG1}, when a vertical current of $j=100$ MA/cm$^{2}$ is injected into the central circular region with a diameter of $30$ nm, the N{\'e}el vector is continuously flipped and then a bimeron-like magnetic structure is formed. At $t=0.05$ ns, the current is turned off.
Since the DMI energy density of the lower half of the magnetic texture has a positive value~\cite{Shen_SI}, the lower half of the magnetic texture is unfavorable and gradually recovers to the AFM ground state, while the upper half evolves into a metastable bimeron.
The current-induced process from the AFM ground state to the metastable bimeron takes only tens of picoseconds, as shown in Fig.~\ref{FIG1}. Such an ultrafast process also exists in the generation of the AFM skyrmions under the action of time-dependent magnetic fields~\cite{Khoshlahni_PRB2019}, where the force induced by time-dependent magnetic fields has a similar form to that of damping-like spin torques~\cite{Tveten_PRL2013,Gomonay_APL2016}. Similar to the AFM skyrmion, the AFM bimeron is a topologically protected magnetic texture with AFM topological charge $Q=\pm 1$, [see Fig.~\ref{FIG1}(i)] where the topological charge is defined as $Q=-\frac{1} {4\pi}\int{dxdy}[\boldsymbol{n}\cdot(\partial_{x}\boldsymbol{n}\times\partial_{y}\boldsymbol{n})]$~\cite{Barker_PRL2016,Lin_PRB2015,Tretiakov_PRB2007}. On the other hand, when the opposite DMI constant is adopted, the AFM bimeron is created in the lower plane (the result is given in Ref.~\cite{Shen_SI}). In addition, for the creation of the AFM bimeron, increasing the injected region can effectively reduce the time and current density, and multiple bimerons will be generated when a small damping is adopted (see Ref.~\cite{Shen_SI}).
Note that the bimeron created here is symmetric, while it may deform under the effect of thermal fluctuations~\cite{Shen_SI}.
%

\begin{figure}[t]
\centerline{\includegraphics[width=0.48\textwidth]{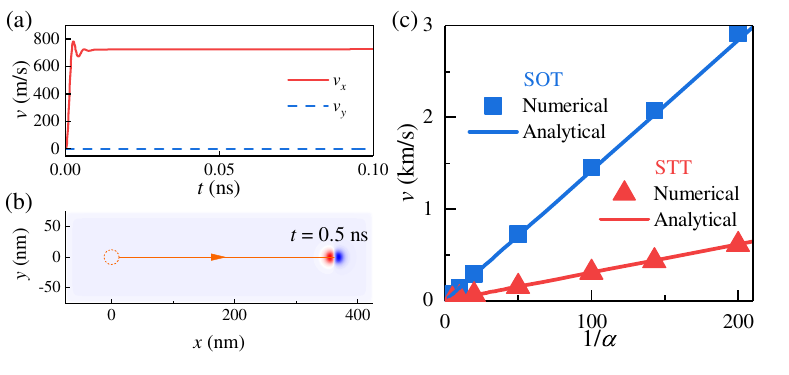}}
\caption{(a) The evolution of the motion speed and (b) the top-view for an AFM bimeron induced by the current via SOTs, where the polarization vector $\boldsymbol{p}= -\boldsymbol{e_{y}}$, the applied current density $j=5$ MA/cm$^{2}$ and the damping $\alpha=0.02$. (c) The motion speed as a function of $1/\alpha$ for the AFM bimeron driven by the current $j=5$ MA/cm$^{2}$ via SOTs and STTs. Symbols are the results of the numerical simulations and lines are given by Eq.~(\ref{eq:4}) with the numerical values of $d \sim 15$ and $R_{s} \sim 7$ nm.}
\label{FIG2}
\end{figure}

\textit{Current-induced motion of an AFM bimeron.}~Manipulating magnetic textures is indispensable in information storage and logic devices. The current, which is a common method to manipulate magnetic materials, is employed to drive the AFM bimeron via SOTs and STTs. Taking the current density $j=5$ MA/cm$^{2}$ and the damping $\alpha=0.02$, we simulate the motion of an AFM bimeron, where the initial state is a metastable AFM bimeron. In order to track the AFM bimeron, the guiding center ($r_{x}$, $r_{y}$) of the bimeron is defined, described as
\begin{equation}
r_{i}=\frac {\int{dxdy}[i\boldsymbol{n}\cdot(\partial_{x}\boldsymbol{n}\times\partial_{y}\boldsymbol{n})]} {\int{dxdy}[\boldsymbol{n}\cdot(\partial_{x}\boldsymbol{n}\times\partial_{y}\boldsymbol{n})]}, i = x,y,\tag{3}
\label{eq:3}
\end{equation}
and the velocity $v_{i} = \dot{r}_{i}$. As shown in Figs.~\ref{FIG2}(a) and (b), considering the damping-like SOTs, the steady motion speed reaches 725 m/s at $t=0.1$ ns and the transmission path of the AFM bimeron is parallel to the racetrack, so that the fast-moving AFM bimeron will not be destroyed by touching the racetrack edge due to the cancellation of the Magnus force. Therefore, in addition to the AFM skyrmions, the AFM bimerons are also ideal information carriers in racetrack-type memory.

Figure~\ref{FIG2}(c) shows the relation between the speed $v$ and the damping $\alpha$, where the speed of the AFM bimeron is inversely proportional to the damping constant for SOTs and STTs. In order to test the simulated speeds, we derived the steady motion speed from Eq.~(\ref{eq:2}) (see Ref.~\cite{Shen_SI} for details)
\begin{equation}
v=\frac {\pi^{2}R_{s}\gamma H_{d}} {\alpha d}-\gamma\frac {\beta} {\alpha}, \tag{4}
\label{eq:4}
\end{equation}
where $R_{s}$ is the bimeron radius, which corresponds to the skyrmion radius.
The first and second terms on the right side of Eq.~(\ref{eq:4}) are the SOT- and STT-induced speeds, respectively.
We can see from Fig.~\ref{FIG2}(c) that the analytical speed given by Eq.~(\ref{eq:4}) is in good agreement with the results of the numerical simulations. It is worth mentioning that Eq.~(\ref{eq:4}) is also applicable to AFM skyrmions. Namely, the AFM bimeron and skyrmion have the same motion speed under the same driving force.

\begin{figure}[t]
\centerline{\includegraphics[width=0.48\textwidth]{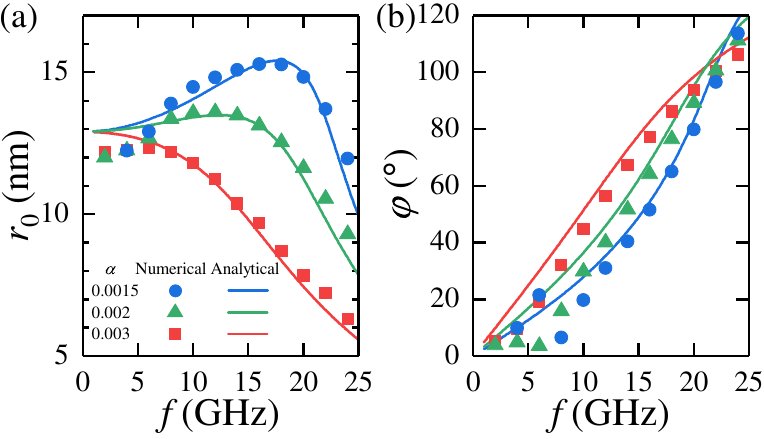}}
\caption{(a) The amplitude $r_{0}$ and (b) phase $\varphi$ as functions of the frequency $f$ of the alternating current [$j = j_{0}$sin($2\pi ft$)], where the symbols are the results of our numerical simulations, while the lines are obtained from Eqs.~(\ref{eq:6}) and (\ref{eq:7}).}
\label{FIG3}
\end{figure}

\textit{Dynamics of the AFM bimeron induced by the alternating current.}~Next, we discuss the forced oscillation of the AFM bimeron induced by the alternating current $j = j_{0}$sin($2\pi ft$), where $j_{0}$ and $f$ are the amplitude and frequency of the applied currents. As shown in Ref.~\cite{Shen_SI}, due to the harmonic current-induced driving force, the guiding center $r_{x}$ of the AFM bimeron exhibits a stable oscillation with amplitude $r_{0} \sim 11.64$ nm and phase difference $\varphi \sim 89.14^{\circ}$ between $j$ and $r_{x}$, where $\alpha = 0.002$, $f$ = 20 GHz and $j_{0}=1$ MA/cm$^{2}$ are adopted. By changing the frequency of the applied currents, the different values of $r_{0}$ and $\varphi$ are obtained by numerical simulations and are shown in Fig.~\ref{FIG3}, where three damping constants ($\alpha = 0.0015$, $0.002$ and $0.003$) are considered. We can see that the phase difference $\varphi$ becomes larger with the increasing frequency, and interestingly, for the amplitude $r_{0}$, there are current-induced resonance phenomena. To analyze such resonance phenomena, we return to Eq.~(\ref{eq:2}) and focus on the motion in the $x$ direction, so that the Thiele equation becomes a scalar equation
\begin{equation}
M_\text{eff}\ddot{r}_{x}+\alpha^{*}\dot{r}_{x}+F_{b}=F_{\text{SOT},0}\text{sin}(2\pi ft), \tag{5}
\label{eq:5}
\end{equation}
where $\alpha^{*}=\alpha\mu_{\text{0}}M_\text{S}t_{z}d/\gamma$ and $F_{\text{SOT},0}\text{sin}(2\pi ft)$ is the force induced by the alternating current with $F_{\text{SOT},0}\approx\pi^{2}R_{s}\mu_{\text{0}}H_{d}M_\text{S}t_{z}$. $F_{b}$ is the boundary-induced force, which can be described as $F_{b}\approx k_{1}r_{x}+k_{2}r^{3}_{x}$ with $k_{1}=4.55\times10^{-6}$ N/m and $k_{2}=2\times10^{10}$ N/m$^3$ for the nanodisk with a diameter of $80$ nm studied here (see Ref.~\cite{Shen_SI} for details). Note that, for other nanodisks, the form of $F_{b}$ may change, resulting in other types of AFM-bimeron-based nonlinear oscillators,

Since $F_{b}$ contains a cubic term, Eq.~(\ref{eq:5}) is called the Duffing equation~\cite{Novak_PRA1982,Moon_SciRep2014}, which describes a nonlinear system. Therefore, the AFM bimeron can be used as a Duffing oscillator, which is promising for various applications, such as in weak signal detection~\cite{Almog_PRL2007,Wang_IEEE1999}. We assume that the solution of Eq.~(\ref{eq:5}) satisfies this form $r_{x}\approx r_{0}\text{sin}(2\pi ft-\varphi)$, and then substituting it into Eq.~(\ref{eq:5}) gives the amplitude $r_{0}$ as
\begin{equation}
r_{0}=\frac {F_{\text{SOT},0}} {\sqrt{[k_{1}+(3/4)k_{2}r^{2}_{0}-M_\text{eff}(2\pi f)^{2}]^{2}+(2\pi f\alpha^{*})^{2}}}, \tag{6}
\label{eq:6}
\end{equation}
and the phase $\varphi$
\begin{equation}
\text{tan}\varphi=\frac {2\pi f\alpha^{*}} {k_{1}+(3/4)k_{2}r^{2}_{0}-M_\text{eff}(2\pi f)^{2}}, \tag{7}
\label{eq:7}
\end{equation}
where $\text{sin}^{3}(2\pi ft-\varphi)\approx (3/4)\text{sin}(2\pi ft-\varphi)$ has been used.
As shown in Fig.~\ref{FIG3}, the results given by Eqs.~(\ref{eq:6}) and (\ref{eq:7}) are consistent with the numerical simulations for all damping constants. We can see from Eqs.~(\ref{eq:6}) and (\ref{eq:7}) that the frequency response depends on the physical quantities of antiferromagnets, such as the damping and the effective mass, so that they may be measured by applying alternating currents. It should be noted that, due to the existence of the nonlinear term ($k_{2}r^{3}_{x}$), Eq.~(\ref{eq:6}) indicates that an alternating current may induce multiple values of $r_{0}$, resulting in the frequency response showing a jump phenomenon. For the nonlinear oscillator based on other types of AFM textures, such as AFM skyrmion and domain wall, one can obtain a similar frequency response. If the nonlinear term and the damping are small, from Eq.~(\ref{eq:6}), the resonance frequency is given by, $f_{r}=1/(2\pi)\sqrt{k_{1}/M_\text{eff}}$, which equals to $16$ GHz for the parameters used here. On the other hand, as mentioned earlier, $r_{0} \sim 11.64$ nm and $\varphi \sim 89.14^{\circ}$ for $f$ = 20 GHz. Eq.~(\ref{eq:7}) indicates that when $f$ is equal to $f_{\pi/2}=1/(2\pi)\sqrt{(k_{1}+0.75k_{2}r^{2}_{0})/M_\text{eff}}$, $\varphi = 90^{\circ}$. Taking $r_{0} = 11.64$ nm, $f_{\pi/2} = 19.2$ GHz is obtained, which is consistent with the simulation result. In addition, for the case of $k_{1} = 0$, $k_{2} = 0$ and $M_\text{eff} = 0$, i.e., there are no boundary effect and effective mass, Eq.~(\ref{eq:7}) also gives the phase $\varphi = 90^{\circ}$, which is independent of the damping and the frequency.

\begin{figure}[t]
\centerline{\includegraphics[width=0.48\textwidth]{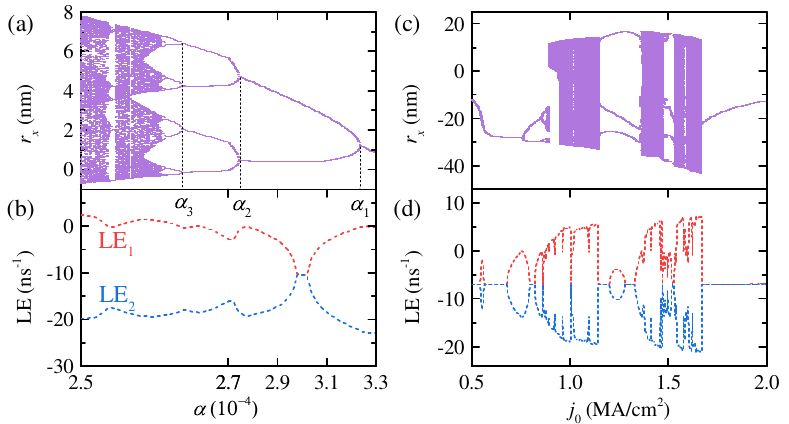}}
\caption{(a) Calculated bifurcation diagram and (b) Lyapunov exponents (LEs) as functions of the damping constant $\alpha$, where $\alpha_{1,2,3}$ = 0.0003236, 0.0002744 and 0.0002638. (c) Bifurcation diagram and (d) LEs as functions of the current density $j$.}
\label{FIG4}
\end{figure}

For a nonlinear system, taking certain parameter values, it shows chaotic behavior. The Lyapunov exponents (LEs) are usually used to judge whether there is chaos, given as~\cite{Yang_PRL2007,Souza-Machado_AmJP1990} 
\begin{equation}
\text{LE}_{i}=\lim\limits_{t \to \infty} \frac{1} {t} \text{ln} \frac{\left \| \delta x^{i}_{t} \right \|} {\left \| \delta x^{i}_{0} \right \|}, \tag{8}
\label{eq:8}
\end{equation}
where $\left \| \delta x^{i}_{0} \right \|$ is the distance between two close trajectories at initial time, and $\left \| \delta x^{i}_{t} \right \|$ is the distance between the trajectories at time $t$. If the largest LE is positive, it means that two close trajectories will be separated. Therefore, a small initial error will increase rapidly, resulting in the evolution of $r_{x}$ being sensitive to initial conditions and its value cannot be predicted for a long time, i.e., the AFM bimeron shows chaotic behavior. Based on Eq.~(\ref{eq:5}), we calculate the bifurcation diagram and LEs (see Ref.~\cite{Shen_SI} for details), and the results are given in Fig.~\ref{FIG4}, showing that the periodic and chaotic windows appear at intervals. We find that a small damping $\alpha$ can lead to the chaotic behavior. The sum of LEs, which equals to $-\alpha\lambda\gamma/\mu_{0} M_{\text{S}}$, agrees with the above result. On the other hand, the value of the damping $\alpha$ at the $i^{\text{th}}$ period-doubling bifurcation should satisfy the universal equation, i.e., the Feigenbaum constant $\delta =\lim\limits_{i \to \infty}[(\alpha_{i}-\alpha_{i-1})/(\alpha_{i+1}-\alpha_{i})] = 4.669$.~\cite{Yang_PRL2007,Souza-Machado_AmJP1990} For the case of Fig.~\ref{FIG4}(a), $\delta_{2}$ is equal to 4.64, from which we estimate that chaos will occur at $\alpha_{\infty} = 0.0002609$.
In addition, the current density $j$ is also of great importance to induce the occurrence of the chaos, as it can be easily tuned in experiment. Figures~\ref{FIG4}(c) and (d) show that for small currents, the system exhibits a periodic movement. With increasing currents, the period-doubling phenomenon takes place and then the system shows chaotic behavior.
It is worth mentioning that the chaotic behavior studied here is subject to the boundary-induced force $F_{b}$, which depends on both the geometric and magnetic parameters. The effects of $F_{b}$ on chaos are discussed in Ref.~\cite{Shen_SI}.


\textit{Conclusions.}
In summary, we have studied the dynamics of an isolated AFM bimeron induced by spin currents. We demonstrate that a spin current can create an isolated bimeron in the AFM film, and drive the AFM bimeron at a speed of a few kilometers per second. Based on the Thiele approach, the steady motion speed is derived, which is in good agreement with the simulation results. Also, we find that the AFM bimeron can be used as the Duffing oscillator. Furthermore, we study the chaotic behavior by calculating the Lyapunov exponents. Our results are useful for the understanding of bimeron physics in AFM systems and may provide guidelines for building spintronic devices based on bimerons.


\textit{Acknowledgments.~}
X.Z. acknowledges the support by the Presidential Postdoctoral Fellowship of The Chinese University of Hong Kong, Shenzhen (CUHKSZ).
M.E. acknowledges the support by the Grants-in-Aid for Scientific Research from JSPS KAKENHI (Grant Nos. JP18H03676, JP17K05490, and JP15H05854) and also the support by CREST, JST (Grant Nos. JPMJCR16F1 and JPMJCR1874).
O.\,A.\,T. acknowledges support by the Cooperative Research Project Program at the Research Institute of Electrical Communication, Tohoku University and by UNSW Science International Seed grant.
X.L. acknowledges the support by the Grants-in-Aid for Scientific Research from JSPS KAKENHI (Grant Nos. 17K19074, 26600041 and 22360122).
G.Z. acknowledges the support by the National Natural Science Foundation of China (Grant Nos. 51771127, 51571126 and 51772004) of China, the Scientific Research Fund of Sichuan Provincial Education Department (Grant Nos. 18TD0010 and 16CZ0006).
Y.Z. acknowledges the support by the President's Fund of CUHKSZ, Longgang Key Laboratory of Applied Spintronics, National Natural Science Foundation of China (Grant Nos. 11974298 and 61961136006), Shenzhen Fundamental Research Fund (Grant No. JCYJ20170410171958839), and Shenzhen Peacock Group Plan (Grant No. KQTD20180413181702403).



\end{document}